# Author's Accepted Manuscript (Post Peer Review Final Draft)

# Deep Neural Network Improves the Estimation of Polygenic Risk Scores for Breast Cancer

Adrien Badré[1], Li Zhang[2], Wellington Muchero[3], Justin C. Reynolds[1], Chongle Pan[1,2,*]

[1]School of Computer Science, University of Oklahoma, Norman, OK
[2]Department of Microbiology and Plant Biology, University of Oklahoma, Norman, OK
[3]BioScience Division, Oak Ridge National Laboratory, Oak Ridge, TN

*Correspondence: cpan@ou.edu

**Abstract**

Polygenic risk scores (PRS) estimate the genetic risk of an individual for a complex disease based on many genetic variants across the whole genome. In this study, we compared a series of computational models for estimation of breast cancer PRS. A deep neural network (DNN) was found to outperform alternative machine learning techniques and established statistical algorithms, including BLUP, BayesA and LDpred. In the test cohort with 50% prevalence, the Area Under the receiver operating characteristic Curve (AUC) were 67.4% for DNN, 64.2% for BLUP, 64.5% for BayesA, and 62.4% for LDpred. BLUP, BayesA, and LPpred all generated PRS that followed a normal distribution in the case population. However, the PRS generated by DNN in the case population followed a bi-modal distribution composed of two normal distributions with distinctly different means. This suggests that DNN was able to separate the case population into a high-genetic-risk case sub-population with an average PRS significantly higher than the control population and a normal-genetic-risk case sub-population with an average PRS similar to the control population. This allowed DNN to achieve 18.8% recall at 90% precision in the test cohort with 50% prevalence, which can be extrapolated to 65.4% recall at 20% precision in a general population with 12% prevalence. Interpretation of the DNN model identified salient variants that were assigned insignificant p-values by association studies, but were important for DNN prediction. These variants may be associated with the phenotype through non-linear relationships.

**Introduction**

Breast cancer is the second deadliest cancer for U.S. women. Approximately one in eight women in the U.S. will develop invasive breast cancer over the course of their lifetime [1]. Early detection of breast cancer is an effective strategy to reduce the death rate. If breast cancer is detected in the localized stage, the 5-year survival rate is 99% [1]. However, only ~62% of the breast cancer cases are detected in the localized stage [1]. In ~30% of the cases, breast cancer is detected after it spreads to the regional lymph nodes, reducing the 5-year survival rate to 85%. Furthermore, in 6% of cases, the cancer is diagnosed after it has spread to a distant part of the body beyond the lymph nodes and the 5-year survival rate is reduced to 27%. To detect breast cancer early, the US Preventive Services Task Force (USPSTF) recommends a biennial screening mammography for women over 50 years old. For women under 50 years old, the decision for screening must be individualized to balance the benefit of potential early detection against the risk of false positive diagnosis. False-positive mammography results, which typically lead to unnecessary follow-up diagnostic testing, become increasingly common for women 40 to 49 years old [2]. Nevertheless, for women with high risk for breast cancer (i.e. a lifetime risk of breast cancer higher than 20%), the American Cancer Society advises a yearly breast MRI and mammogram starting at 30 years of age [3].

Polygenic risk scores (PRS) assess the genetic risks of complex diseases based on the aggregate statistical correlation of a disease outcome with many genetic variations over the whole genome. Single-nucleotide polymorphisms (SNPs) are the most commonly used genetic variations. While genome-wide association studies (GWAS) report only SNPs with statistically significant associations to phenotypes [4], PRS can be estimated using a greater number of SNPs with higher adjusted p-value thresholds to improve prediction accuracy.

Previous research has developed a variety of PRS estimation models based on Best Linear Unbiased Prediction (BLUP), including gBLUP [5] , rr-BLUP [6], [7], and other derivatives [8], [9]. These linear mixed models consider genetic variations as fixed effects and use random effects to account for environmental factors and individual variability. Furthermore, linkage disequilibrium was utilized as a basis for the LDpred [10], [11] and PRS-CS [12] algorithms

PRS estimation can also be defined as a supervised classification problem. The input features are genetic variations and the output response is the disease outcome. Thus, machine learning techniques can be used to estimate PRS based on the classification scores achieved [13]. A large-scale GWAS dataset may provide tens of thousands of individuals as training examples for model development and benchmarking. Wei et al (2019) [14] compared support vector machine and logistic regression to estimate PRS of Type-1 diabetes. The best Area Under the receiver operating characteristic Curve (AUC) was 84% in this study. More recently, neural networks have been used to estimate human height from the GWAS data, and the best $R^2$ scores were in the range of 0.4 to 0.5 [15]. Amyotrophic lateral sclerosis was also investigated using Convolutional Neural Networks (CNN) with 4511 cases and 6127 controls [16] and the highest accuracy was 76.9%.

Significant progress has been made for estimating PRS for breast cancer from a variety of populations. In a recent study [17], multiple large European women cohorts were combined to compare a series of PRS models. The most predictive model in this study used lasso regression with 3,820 SNPs and obtained an AUC of 65%. A PRS algorithm based on the sum of log odds ratios of important SNPs for breast cancer was used in the Singapore Chinese Health Study [18] with 46 SNPs and 56.6% AUC, the Shanghai Genome-Wide Association Studies [19] with 44 SNPs and 60.6% AUC, and a Taiwanese cohort [20] with 6 SNPs and 59.8% AUC. A pruning and thresholding method using 5,218 SNPs reached an AUC of 69% for the UK Biobank dataset [11].

In this study, deep neural network (DNN) was tested for breast cancer PRS estimation using a large cohort containing 26053 cases and 23058 controls. The performance of DNN was shown to be higher than alternative machine learning algorithms and other statistical methods in this large cohort. Furthermore, DeepLift [21] and LIME [22] were used to identify salient SNPs used by DNN for prediction.

**Materials and Methods**

Breast cancer GWAS data

This study used a breast cancer GWAS dataset generated by the Discovery, Biology, and Risk of Inherited Variants in Breast Cancer (DRIVE) project [23] and was obtained from the NIH dbGaP database under the accession number of phs001265.v1.p1. The DRIVE dataset was stored, processed and used on the Schooner supercomputer at the University of Oklahoma in an isolated partition with restricted access. The partition consisted of 5 computational nodes, each with 40 CPU cores (Intel Xeon Cascade Lake) and 200 GB of RAM. The DRIVE dataset in the dbGap database was composed of 49,111 subjects genotyped for 528,620 SNPs using OncoArray [23]. 55.4% of the subjects were from North America, 43.3% from Europe, and 1.3% from Africa. The disease outcome of the subjects was labeled as malignant tumor (48%), *in situ* tumor (5%), and no tumor (47%). In this study, the subjects in the malignant tumor and *in situ* tumor categories were labeled as cases and the subjects in the no tumor category were labeled as controls, resulting in 26053 (53%) cases and 23058 (47%) controls. The subjects in the case and control classes were randomly assigned to a training set (80%), a validation set (10%), and a test set (10%) (Figure 1). The association analysis was conducted on the training set using Plink 2.0 [24]. For a subject, each of the 528,620 SNPs may take the value of 0,1 or 2, representing the genotype value on a SNP for this subject. The value of 0 meant homozygous with minor allele, 1 meant heterozygous allele, and 2 meant homozygous with the dominant allele. Such encoding of the SNP information was also used in the following machine learning and statistical approaches. The p-value for each SNP was calculated using logistic regression in Plink 2.0.

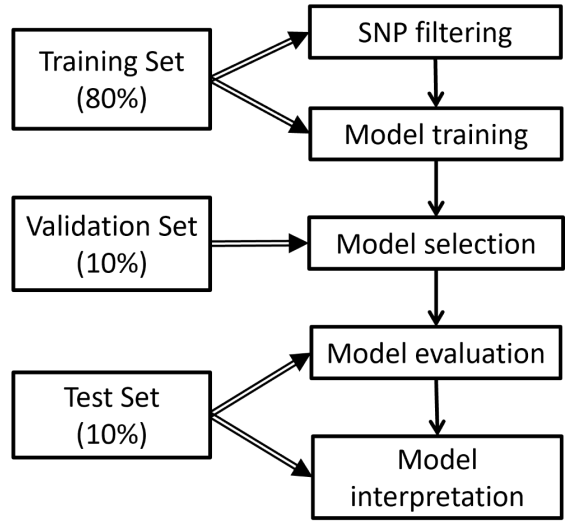

**Figure 1: Computational workflow of predictive genomics.** The DRIVE dataset was randomly split into the training set, the validation set, and the test set. Only the training set was used for association analysis, which generated the p-values for selection of SNPs as input features. The training data was then used to train machine learning models and statistical models. The validation set was used to select the best hyperparameters for each model based on the validation AUC score. Finally, the test set was used for performance benchmarking and model interpretation.

Development of deep neural network models for PRS estimation

A variety of deep neural network (DNN) architectures [25] were trained using Tensorflow 1.13. The Leaky Rectified Linear Unit (ReLU) activation function [26] was used on all hidden-layers neurons with the negative slope co-efficient set to 0.2. The output neuron used a sigmoid activation function. The training error was computed using the cross-entropy function:

$$\sum_{i=1}^{n} y * \log(p) + (1 - y) * \log(1 - p),$$

where $p \in [0,1]$ is the prediction probability from the model and $y \in [\![0,1]\!]$ is the prediction target at 1 for case and 0 for control. The prediction probability was considered as the PRS from DNN.

DNNs were evaluated using different SNP feature sets. SNPs were filtered using their Plink association p-values at the thresholds of $10^{-2}, 10^{-3}, 10^{-4}$ and $10^{-5}$. DNN was also tested using the full SNP feature set without any filtering. The DNN models were trained using mini-batches with a batch size of 512. The Adam optimizer [27], an adaptive learning rate optimization algorithm, was used to update the weights in each mini-batch. The initial learning rate was set to $10^{-4}$, and the models were trained for up to 200 epochs with early stopping based on the validation AUC score. Dropout [28] was used to reduce overfitting. The dropout rates of 20%, 30%, 40%, 50%, 60%, 70%, 80%, and 90% were tested for the first hidden layer and the final dropout rate was selected based on the validation AUC score. The dropout rate was set to 50% on the other hidden layers in all architectures. Batch normalization (BN) [29] was used to accelerate the training process, and the momentum for the moving average was set to 0.9 in BN.

Development of alternative machine learning models for PRS estimation

Logistic regression, decision tree, random forest, AdaBoost, gradient boosting, support vector machine (SVM), and Gaussian naive Bayes were implemented and tested using the scikit-learn machine learning library in Python. These models were trained using the same training set as the DNNs and their hyperparameters were tuned using the same validation set based on the validation AUC (Figure 1). These models are briefly described below.

- **Decision Tree:** The gini information gain with best split was used. The maximum depth was not set to let the tree expanded until all leaves were pure or contained less than a minimum number of two examples per split (sklearn default parameters).
- **Random Forest:** classification decision trees (as configured above) were used as base learners. The optimum number of decision trees were found to be 3,000 based on a parameter sweep between 500 and 5,000 with a step size of 500. Bootstrap samples were used to build each base learner. When searching for each tree's best split, the maximum number of considered features was set to be the square root of the number of features.
- **AdaBoost**: classification decision trees (as configured above) were used as base learners. The optimum number of decision trees were found to be 2,000 based on a parameter sweep between 500 and 5,000 with a step size of 500. The learning rate was set to 1. The algorithm used was SAMME.R [30].
- **Gradient Boosting:** regression decision trees (as configured above) were used as the base learners. The optimum number of decision trees were found to be 400 based on a parameter sweep between 100 and 1,000 with a step size of 100. Log-loss was used as the loss function. The learning rate was set to 0.1. The mean squared error with improvement score [31] was used to measure the quality of a split.

- **SVM:** The kernel was a radial basis function with $\gamma = \frac{1}{n*Var}$, where $n$ is the number of SNPs and $Var$ is the variance of the SNPs across individuals. The regularization parameter C was set to 1 based on a parameter sweep over 0.001, 0.01, 0.1, 1, 5, 10, 15 and 20.
- **Logistic Regression:** L2 regularization with $\alpha = 0.5$ was used based on a parameter sweep for $\alpha$ over 0.0001, 0.001, 0.01, 0.1, 0.2, 0.3, 0.4, 0.5, 0.6, 0.7 and 0.8. L1 regularization was tested, but not used, because it did not improve the performance.
- **Gaussian Naïve Bayes:** The likelihood of the features was assumed to be Gaussian. The classes had uninformative priors.

Development of statistical models for PRS estimation

The same training and validation sets were used to develop statistical models (Figure 1). The BLUP and BayesA models were constructed using the bWGR R package. The LDpred model was constructed as described [10].

- **BLUP:** The linear mixed model was $y = \mu + Xb + e$, where y were the response variables, $\mu$ were the intercepts, X were the input features, b were the regression coefficients, and e were the residual coefficients.
- **BayesA:** The priors were assigned from a mixture of normal distributions.
- **LDpred**: The p-values were generated by our association analysis described above. The validation set was provided as reference for LDpred data coordination. The radius of the Gibbs sampler was set to be the number of SNPs divided by 3000 as recommended by the LDpred user manual (https://github.com/bvilhjal/ldpred/blob/master/ldpred/run.py).

The score distributions of DNN, BayesA, BLUP and LDpred were analyzed with the Shapiro test for normality and the Bayesian Gaussian Mixture (BGM) expectation maximization algorithm. The BGM algorithm decomposed a mixture of two Gaussian distributions with weight priors at 50% over a maximum of 1000 iterations and 100 initializations.

DNN model interpretation.

LIME and DeepLift were used to interpret the DNN predictions for subjects in the test set with DNN output scores higher than 0.67, which corresponded to a precision of 90%. For LIME, the submodular pick algorithm was used, the kernel size was set to 40, and the number of explainable features was set to 41. For DeepLift, the importance of each SNPs was computed as the average across all individuals, and the reference activation value for a neuron was determined by the average value of all activations triggered across all subjects.

**Results and Discussion**

Development of a machine learning model for breast cancer PRS estimation

The breast cancer GWAS dataset containing 26053 cases and 23058 controls was generated by the Discovery, Biology, and Risk of Inherited Variants in Breast Cancer (DRIVE) project [23]. The DRIVE data is available from the NIH dbGaP database under the accession number of phs001265.v1.p1. The cases and controls were randomly split to a training set, a validation set, and a test set (Figure 1). The training set was used to estimate p-values of SNPs using association analysis and train machine learning and statistical models. The hyperparameters of the machine learning and statistical models were optimized using the validation set. The test set was used for the final performance evaluation and model interpretation.

Statistical significance of the disease association with the 528,620 SNPs was assessed with Plink using only the training set. To obtain unbiased benchmarking results on the test set, it was critical not to use the test set in the association analysis (Figure 1) and not to use association p-values from previous GWAS studies that included subjects in the test set, as well-described in the Section 7.10.2 of Hastie et al [32]. The obtained p-values for all SNPs are shown in Figure 2A as a Manhattan plot. There were 1,061 SNPs with a p-value less than the critical value of $9.5 \cdot 10^{-8}$, which was set using the Bonferroni correction ($9.5 \cdot 10^{-8} = 0.05/528{,}620$). Filtering with a Bonferroni-corrected critical value may remove many informative SNPs that have small effects on the phenotype, epistatic interactions with other SNPs, or non-linear association with the phenotype [33]. Relaxed filtering with higher p-value cutoffs was tested to find the optimal feature set for DNN (Figure 2B and Supplementary Table 1). The DNN models in Figure 2B had a deep feedforward architecture consisting of an input layer of variable sizes, followed by 3 successive hidden layers containing 1000, 250, and 50 neurons, and finally an output layer with a single neuron. As the p-value cutoff increased, a greater number of SNPs were incorporated as input features, and training consumed a larger amount of computational resources in terms of computing time and peak memory usage. A feature set containing 5,273 SNPs above the p-value cutoff of $10^{-3}$ provided the best prediction performance measured by the AUC and accuracy on the validation set. In comparison with smaller feature sets from more stringent p-value filtering, the 5,273-SNP feature set may have included many informative SNPs providing additional signals to be captured by DNN for prediction. On the other hand, more relaxed filtering with p-value cutoffs greater than $10^{-3}$ led to significant overfitting as indicated by an increasing prediction performance in the training set and a decreasing performance in the validation set (Figure 2B).

    Previous studies [11], [34] have used a large number of SNPs for PRS estimation on different datasets. In our study, the largest DNN model, consisting of all 528,620 SNPs, decreased the validation AUC score by 1.2% and the validation accuracy by 1.9% from the highest achieved values. This large DNN model relied an 80% dropout rate to obtain strong regularization, while all the other DNN models utilized a 50% dropout rate. This suggested that DNN was able to perform feature selection without using association p-values, although the limited training data and the large neural network size resulted in complete overfitting with a 100% training accuracy and the lowest validation accuracy (Figure 2B).

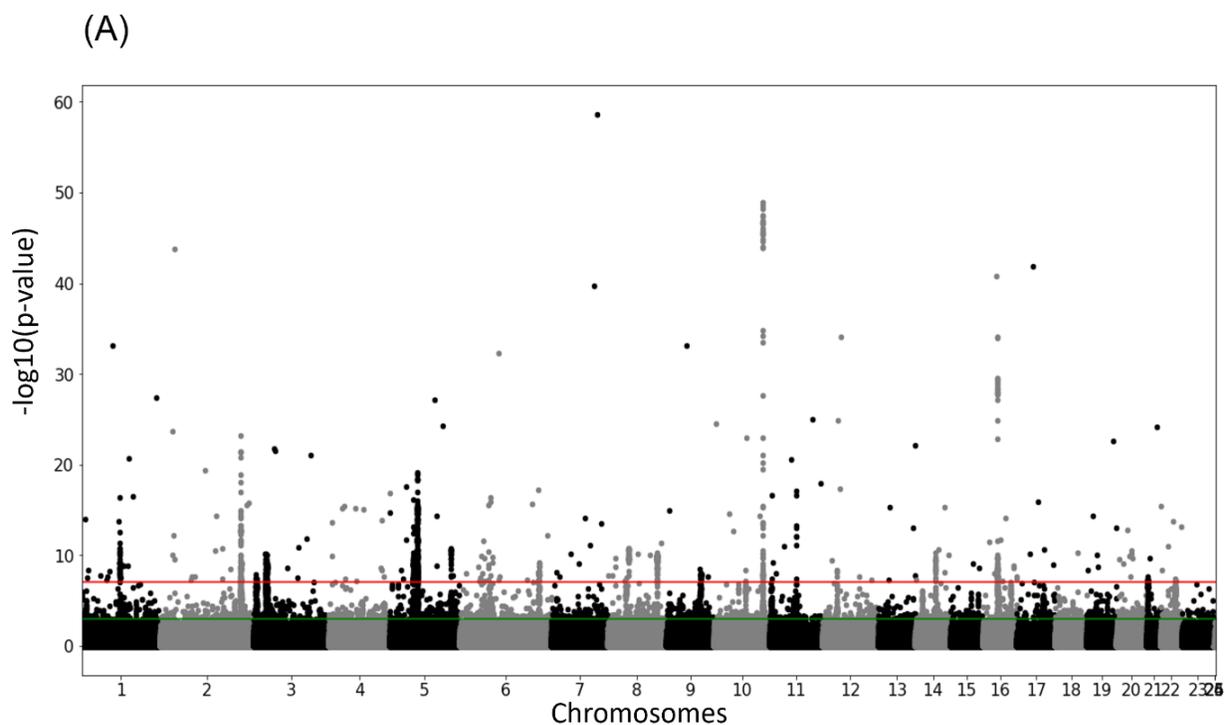

| p-value cutoff | SNPs | Computational Cost of Training | | AUC | | Accuracy | |
|---|---|---|---|---|---|---|---|
| | | Convergence time (minutes) | Peak Memory (GB) | Training | Validation | Training | Validation |
| None | 528,620 | 1308 | 66.6 | 100.0% | 65.9% | 100.0% | 60.1% |
| $10^{-2}$ | 13,890 | 51 | 3.2 | 93.4% | 66.5% | 85.1% | 61.4% |
| $10^{-3}$ | 5,273 | 23 | 2.2 | 80.5% | 67.1% | 73.4% | 62.0% |
| $10^{-4}$ | 3,041 | 16 | 2 | 75.9% | 66.4% | 67.6% | 61.1% |
| $10^{-5}$ | 2,099 | 9 | 1.4 | 72.2% | 65.7% | 63.2% | 60.8% |

**Figure 2: SNP filtering and model training for DNN. (A)** Manhattan plot from the association analysis. Each point represents a SNP with its p-value in the log10 scale on the y-axis and its position in a chromosome on the x-axis. The x-axis is labeled with the chromosome numbers. Chromosome 23 represents the X chromosome. Chromosomes 24 and 25 represent the pseudoautosomal region and non-pseudoautosomal region of the Y chromosome, respectively. Chromosome 26 designates mitochondrial chromosome. The red line marks the p-value cutoff at $9.5 \cdot 10^{-8}$ and the green line marks the p-value cutoff at $10^{-3}$. **(B)** Performance of the DNN models trained using five SNP sets filtered with increasing p-value cutoffs. The models were compared by their training costs and performances in the training and validation sets.

The effects of dropout and batch normalization were tested using the 5,273-SNP DNN model (Supplementary Figure 1). Without dropout, the DNN model using only batch normalization had a 3.0% drop in AUC and a 4.0% drop in accuracy and its training converged in only two epochs. Without batch normalization, the DNN model had 0.1% higher AUC and 0.3% lower accuracy but its training required a 73% increase in the number of epochs to reach convergence.

As an alternative to filtering, autoencoding was tested to reduce SNPs to a smaller set of encodings as described previously [35], [36]. An autoencoder was trained to encode 5273 SNPs into 2000 features with a mean square error (MSE) of 0.053 and a root mean square error (RMSE) of 0.23. The encodings from the autoencoder were used as the input features to train a DNN model with the same architecture as the ones shown in Figure 2B except for the number of input neurons. The autoencoder-DNN model had a similar number of input neurons for DNN as the 2099-SNP DNN model, but had a 1.3% higher validation AUC and a 0.2% higher validation accuracy than the 2099-SNP DNN model (Figure 2B). This increased validation AUC and accuracy suggested that the dimensionality reduction by the autoencoding from 5273 SNPs to 2000 encodings was better than the SNP filtering by the association p-values from 5273 SNPs to 2099 SNPs. However, the DNN models with 5,273 SNPs still had a 0.3% higher validation AUC score and a 1.6% higher validation accuracy than the autoencoder-DNN model.

The deep feedforward architecture benchmarked in Figure 2B was compared with a number of alternative neural network architectures using the 5,273-SNP feature set (Supplementary Table 2). A shallow neural network with only one hidden layer resulted in a 0.9% lower AUC and 1.1% lower accuracy in the validation set compared to the DNN. This suggested that additional hidden layers in DNN may allow additional feature selection and transformation in the model. One-dimensional convolutional neural network (1D CNN) was previously used to estimate the PRS for bone heel mineral density, body mass index, systolic blood pressure and waist-hip ratio [15] and was also tested here for breast cancer prediction with the DRIVE dataset. The validation AUC and accuracy of 1D CNN was lower than DNN by 3.2% and 2.0%, respectively. CNN was commonly used for image analysis, because the receptive field of the convolutional layer can capture space-invariant information with shared parameters. However, the SNPs distributed across a genome may not have significant space-invariant patterns to be captured by the convolutional layer, which may explain the poor performance of CNN.

The 5,273-SNP feature set was used to test alternative machine learning approaches, including logistic regression, decision tree, naive Bayes, random forest, ADAboost, gradient boosting, and SVM, for PRS estimation (Figure 3). These models were trained, turned, and benchmarked using the same training, validation, and test sets, respectively, as the DNN models (Figure 1). Although the decision tree had a test AUC of only 50.9%, ensemble algorithms that used decision trees as the base learner, including random forest, ADABoost, and gradient boosting, reached test AUCs of 63.6%, 64.4%, and 65.1%, respectively. This showed the advantage of ensemble learning. SVM reached a test AUC of 65.6%. Naïve Bayes and logistic regression were both linear models with the assumption of independent features. Logistic regression had higher AUC, but lower accuracy, than SVM and gradient boosting. The test AUC and test accuracy of DNN were higher than those of logistic regression by 0.9% and 2.7%, respectively. Out of all the machine learning models, the DNN model achieved the highest test AUC at 67.4% and the highest test accuracy at 62.8% (Figure 3).

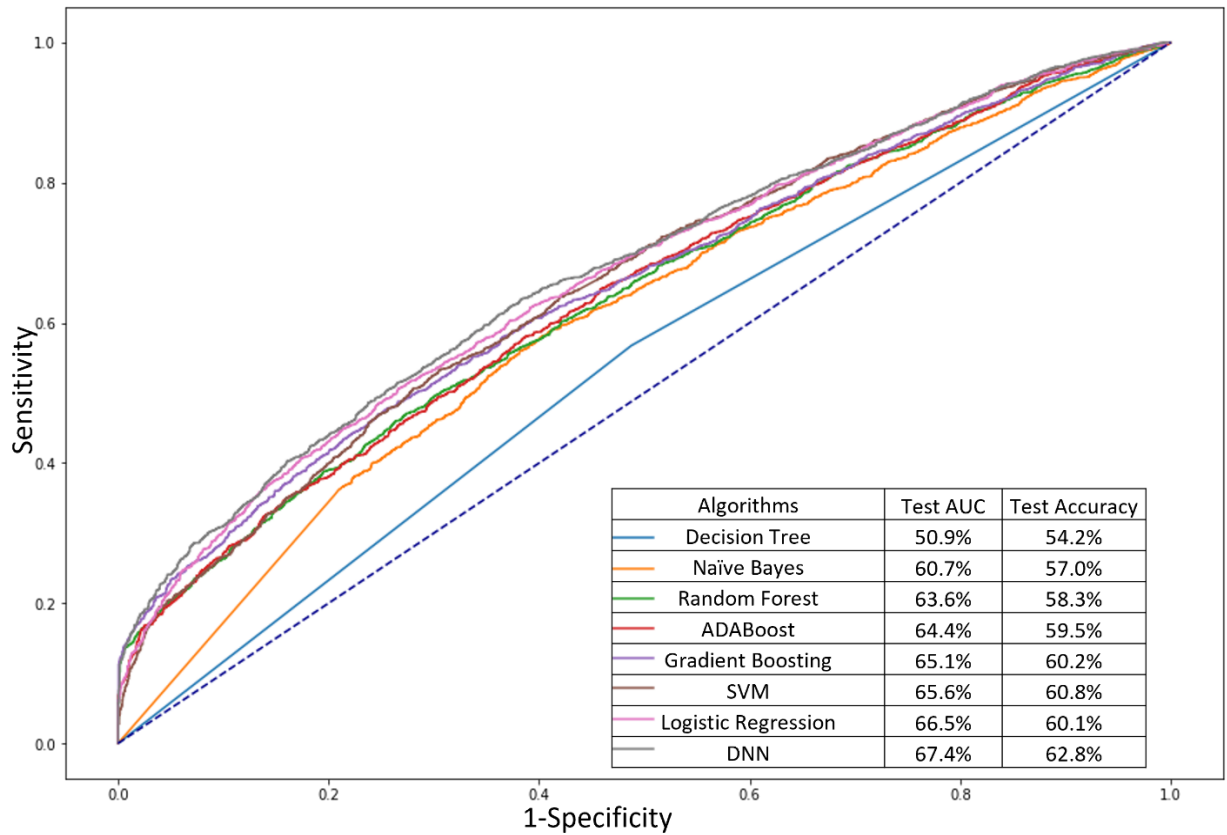

**Figure 3: Comparison of machine learning approaches for PRS estimation.** The performance of the models were represented as Receiver Operating Characteristic (ROC) curves in different colors. The Area under the ROC curve (AUC) and the accuracy from the test set are shown in the legend. The DNN model outperformed the other machine learning models in terms of AUC and accuracy.

Comparison of the DNN model with statistical models for breast cancer PRS estimation

The performance of DNN was compared with three representative statistical models, including BLUP, BayesA, and LDpred (Table 1). Because the relative performance of these methods may be dependent on the number of training examples available, the original training set containing 39,289 subjects was down-sampled to create three smaller training sets containing 10000, 20000, 30000 subjects. As the 5,273-SNP feature set generated with a p-value cutoff of $10^{-3}$ may not be the most appropriate for the statistical methods, a 13,890-SNP feature set (p-value cutoff = $10^{-2}$) and a 2,099-SNP feature set (p-value cutoff = $10^{-5}$) were tested for all methods.

**Table 1. AUC test scores of DNN, BLUP, BayesA and LDpred models at different p-value cutoffs (PC) and training set sizes (TS).**

| Algorithms | DNN | | | BLUP | | | BayesA | | | LDpred | | |
|---|---|---|---|---|---|---|---|---|---|---|---|---|
| PC* \ TS** | $10^{-5}$ | $10^{-3}$ | $10^{-2}$ | $10^{-5}$ | $10^{-3}$ | $10^{-2}$ | $10^{-5}$ | $10^{-3}$ | $10^{-2}$ | $10^{-5}$ | $10^{-3}$ | $10^{-2}$ |
| 10,000 | 64.8% | 65.5% | 65.1% | 63.5% | 62.5% | 60.6% | 63.7% | 62.0% | 59.9% | 60.8% | 62.4% | 61.6% |
| 20,000 | 65.6% | 66.6% | 66.4% | 62.9% | 63.0% | 60.6% | 62.7% | 63.0% | 60.4% | 60.8% | 62.4% | 61.6% |
| 30,000 | 66.0% | 66.9% | 66.6% | 64.2% | 63.1% | 60.7% | 64.3% | 63.1% | 60.7% | 60.7% | 62.4% | 61.6% |
| 39,289 | 66.2% | 67.4% | 67.3% | 64.2% | 63.3% | 61.0% | 64.5% | 63.4% | 61.1% | 60.7% | 62.4% | 61.6% |

\*: p-value cutoff
\*\*: training set size

Although LDpred also required training data, its prediction relied primarily on the provided p-values, which were generated for all methods using all 39,289 subjects in the training set. Thus, the down-sampling of the training set did not reduce the performance of LDpred. LDpred reached its highest AUC score at 62.4% using the p-value cutoff of $10^{-3}$. A previous study [12] that applied LDpred to breast cancer prediction using the UK Biobank dataset similarly obtained an AUC score of 62.4% at the p-value cutoff of $10^{-3}$. This showed consistent performance of LDpred in the two studies. When DNN, BLUP, and BayesA used the full training set, they obtained higher AUCs than LDpred at their optimum p-value cutoffs.

DNN, BLUP, and BayesA all gained performance with the increase in the training set sizes (Table 1). The performance gain was more substantial for DNN than BLUP and BayesA. The increase from 10,000 subjects to 39,258 subjects in the training set resulted in a 1.9% boost to DNN's best AUC, a 0.7% boost to BLUP, and a 0.8% boost to BayesA. This indicated the different variance-bias trade-offs made by DNN, BLUP, and BayesA. The high variance of DNN required more training data, but could capture non-linear relationships between the SNPs and the phenotype. The high bias of BLUP and BayesA had lower risk for overfitting using smaller training sets, but their models only considered linear relationships. The higher AUCs of DNN across all training set sizes indicated that DNN had a better variance-bias balance for breast cancer PRS estimation.

For all four training set sizes, BLUP and BayesA achieved higher AUCs using more stringent p-value filtering. When using the full training set, reducing the p-value cutoffs from $10^{-2}$ to $10^{-5}$ increased the AUCs of BLUP from 61.0% to 64.2% and the AUCs of BayesA from 61.1% to 64.5%. This suggested that BLUP and BayesA preferred a reduced number of SNPs that were significantly associated with the phenotype. On the other hand, DNN produced lower AUCs using the p-value cutoff of $10^{-5}$ than the other two higher cutoffs. This suggested that DNN can perform better feature selection in comparison to SNP filtering based on association p-values.

The four algorithms were compared using the PRS histograms of the case population and the control population from the test set in Figure 4. The score distributions of BLUP, BayesA and LDpred all followed normal distributions. The p-values from the Shapiro normality test of the case and control distributions were 0.46 and 0.43 for BayesA, 0.50 and 0.95 for BLUP, and 0.17 and 0,24 for LDpred, respectively. The case and control distributions were $N_{case}(\mu = 0.577, \sigma = 0.20)$ and $N_{control}(\mu = 0.479, \sigma = 0.19)$ from BayesA, $N_{cases}(\mu = 0.572, \sigma = 0.19)$ and $N_{control}(\mu = 0.483, \sigma = 0.18)$ from BLUP, and $N_{case}(\mu = -33.52, \sigma = 5.4)$ and $N_{control}(\mu = -35.86, \sigma = 4.75)$ from LDpred. The means of the case distributions were all significantly higher than the control distributions for BayesA (p-value < $10^{-16}$), BLUP (p-value < $10^{-16}$), and LDpred (p-value < $10^{-16}$) and their case and control distributions had similar standard deviations.

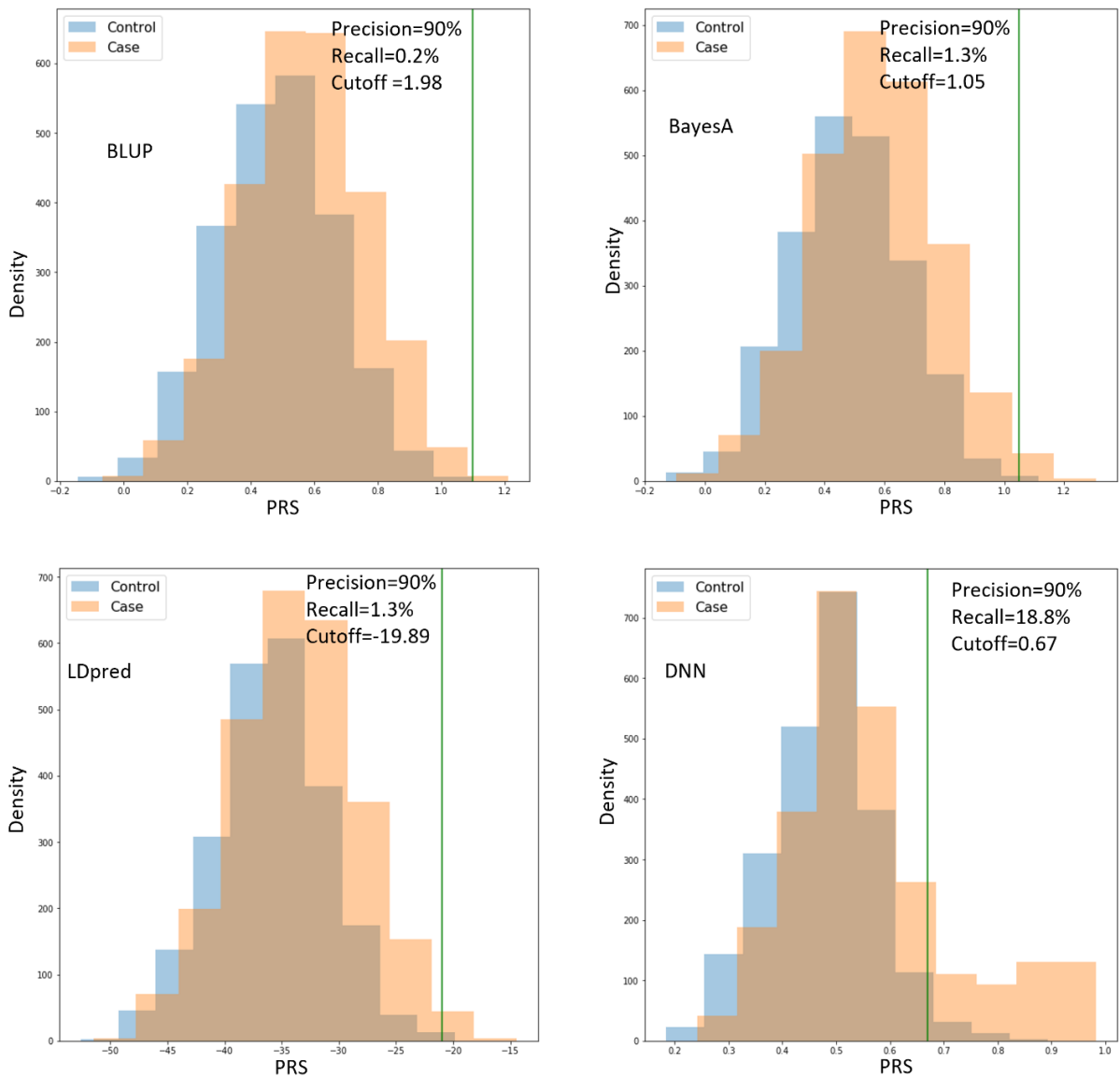

**Figure 4: Score histograms of DNN, BLUP, BayesA and LDpred.** The case and control populations are shown in the orange and blue histograms, respectively. The green line represents the score cutoff corresponding to the precision of 90% for each model. DNN had a much higher recall than the other algorithms at the 90% precision.

The score histograms of DNN did not follow normal distributions based on the Shapiro normality test with a p-value of $4.1 * 10^{-34}$ for the case distribution and a p-value of $2.5 * 10^{-9}$ for the control distribution. The case distribution had the appearance of a bi-modal distribution. The Bayesian Gaussian mixture expectation maximization algorithm decomposed the case distribution to two normal distributions: $N_{case1}(\mu = 0.519, \sigma = 0.096)$ with an 86.5% weight and $N_{case2}(\mu = 0.876, \sigma = 0.065)$ with a 13.5% weight. The control distribution was resolved into two normal distributions with similar means and distinct standard deviations: $N_{control1}(\mu = 0.471, \sigma = 0.1)$ with an 85.0% weight and $N_{control2}(\mu = 0.507, \sigma = 0.03)$ with a 15.0% weight. The $N_{case1}$ distribution had a similar mean as the $N_{control1}$ and $N_{control2}$ distributions. This suggested that the $N_{case1}$ distribution may represent a normal-genetic-risk case sub-population, in which the subjects may have a normal level of genetic risk for breast cancer and the oncogenesis likely involved a significant environmental component. The mean of the $N_{case2}$ distribution was higher than the means of both the $N_{case1}$ and $N_{control1}$ distributions by more than 4 standard deviations (p-value < $10^{-16}$). We hypothesized that the $N_{case2}$ distribution represented a high-genetic-risk case sub-population for breast cancer, in which the subjects may have inherited many genetic variations associated with breast cancer.

Three GWAS were performed between the high-genetic-risk case sub-population with DNN PRS > 0.67, the normal-genetic-risk case sub-population with DNN PRS < 0.67, and the control population (Supplementary Table 3). The GWAS analysis of the high-genetic-risk case sub-population versus the control population identified 182 significant SNPs at the Bonferroni level of statistical significance. The GWAS analysis of the high-genetic-risk case sub-population versus the normal-genetic-risk case sub-population identified 216 significant SNPs. The two sets of significant SNPs found by these two GWAS analyses were very similar, sharing 149 significant SNPs in their intersection. Genes associated with these 149 SNPs were investigated with pathway enrichment analysis (Fisher's Exact Test; P < 0.05) using SNPnexus [37] (Supplementary Table 4). Many of the significant pathways were involved in DNA repair [38] signal transduction [39], and suppression of apoptosis [40]. Interestingly, the GWAS analysis of the normal-genetic-risk case sub-population and the control population identified no significant SNP. This supported our classification of the cases into the normal-genetic-risk subjects and the high-genetic-risk subjects based on their PRS scores from the DNN model.

In comparison with AUCs, it may be more relevant for practical applications of PRS to compare the recalls of different algorithms at a given precision that warrants clinical recommendations. At 90% precision, the recalls were 18.8% for DNN, 0.2% for BLUP, 1.3% for BayesA, and 1.3% for LDpred in the test set of the DRIVE cohort with a ~50% prevalence. This indicated that DNN can make a positive prediction for 18.8% of the subjects in the DRIVE cohort and these positive subjects would have an average chance of 90% to eventually develop breast cancer. American Cancer Society advises yearly breast MRI and mammogram starting at the age of 30 years for women with a lifetime risk of breast cancer greater than 20%, which meant a 20% precision for PRS. By extrapolating the performance in the DRIVE cohort, the DNN model should be able to achieve a recall of 65.4% at a precision of 20% in the general population with a 12% prevalence rate of breast cancer.

Interpretation of the DNN model

While the DNN model used 5,273 SNPs as input, we hypothesized that only a small set of these SNPs were particularly informative for identifying the subjects with high genetic risks for breast cancer. LIME and DeepLift were used to find the top-100 salient SNPs used by the DNN model to identify the subjects with PRS higher than the 0.67 cutoff at 90% precision in the test set (Figure 1). Twenty three SNPs were ranked by both algorithms to be among their top-100 salient SNPs (Supplementary Figure 2). The small overlap between their results can be attributed to their different interpretation approaches. LIME considered the DNN model as a black box and perturbed the input to estimate the importance of each variable; whereas, DeepLift analyzed the gradient information of the DNN model. 30% of LIME's salient SNPs and 49% of DeepLift's salient SNPs had p-values less than the Bonferroni significance threshold of $9.5 \cdot 10^{-8}$. This could be attributed to the non-linear relationships between the salient SNP genotype and the disease outcome, which cannot be captured by the association analysis using logistic regression. To illustrate this, four salient SNPs with significant p-values were shown in Supplementary Figure 3A, which exhibited linear relationships between their genotype values and log odds ratios as expected. Four salient SNPs with insignificant p-values were shown in Supplementary Figure 3B, which showed clear biases towards cases or controls by one of the genotype values in a non-linear fashion.

Michailidiou et al. [41] summarized a total of 172 SNPs associated with breast cancer. Out of these SNPs, 59 were not included on OncoArray, 63 had an association p-value less than $10^{-3}$ and were not included in the 5,273-SNP feature set for DNN, 34 were not ranked among the top-1000 SNPs by either DeepLIFT or LIME, and 16 were ranked among the top-1000 SNPs by DeepLIFT, LIME or both (Supplementary Table 5). This indicates that many SNPs with significant association may be missed by the interpretation of DNN models.

The 23 salient SNPs identified by both DeepLift and LIME in their top-100 list are shown in Table 2. Eight of the 23 SNPs had p-values higher than the Bonferroni level of significance and were missed by the association analysis using Plink. The potential oncogenesis mechanisms for some of the 8 SNPs have been investigated in previous studies. The SNP, rs139337779 at 12q24.22, is located within the gene, Nitric oxide synthase 1 (NOS1). Li et al. [42] showed that the overexpression of NOS1 can up-regulate the expression of ATP-binding cassette, subfamily G, member 2 (ABCG2), which is a breast cancer resistant protein [43], and NOS1-indeuced chemo-resistance was partly mediated by the up-regulation of ABCG2 expression. Lee et al. [44] reported that NOS1 is associated with the breast cancer risk in a Korean cohort. The SNP, chr13_113796587_A_G at 13q34, is located in the F10 gene, which is the coagulation factor X. Tinholt et al [45] showed that the increased coagulation activity and genetic polymorphisms in the F10 gene are associated with breast cancer. The BNC2 gene containing the SNP, chr9_16917672_G_T at 9p22.2, is a putative tumor suppressor gene in high-grade serious ovarian carcinoma [46]. The SNP, chr2_171708059_C_T at 2q31.1, is within the GAD1 gene and the expression level of GAD1 is a significant prognostic factor in lung adenocarcinoma [47]. Thus, the interpretation of DNN models may identify novel SNPs with non-linear association with the breast cancer.

**Table 2: Top salient SNPs identified by both LIME and DeepLift from the DNN model.**

| SNP | locus | LIME ($10^{-4}$) | DeepLift ($10^{-2}$) | p-value | MAF* | Genes of interest** |
|---|---|---|---|---|---|---|
| corect_rs139337779 | 12q24.22 | 4.5 | -3.3 | 6.5E-04 | 11% | NOS1 |
| chr13_113796587_A_G | 13q34 | 4.3 | -3.8 | 2.8E-04 | 3% | F10 |
| chr9_16917672_G_T | 9p22.2 | 4.5 | -2.5 | 7.6E-05 | 4% | BNC2/RP11-132E11.2 |
| chr8_89514784_A_G | 8q21.3 | 27.0 | 3.7 | 2.5E-05 | 56% | RP11-586K2.1 |
| chr17_4961271_G_T | 17p13.2 | 4.2 | -2.2 | 8.2E-06 | 4% | SLC52A1/RP11-46I8.1 |
| rs11642757 | 16q23.2 | 5.3 | -2.9 | 2.0E-06 | 6% | RP11-345M22.1 |
| rs4040605 | 1p36.33 | 4.4 | 2.4 | 9.6E-07 | 37% | RP11-54O7.3/SAMD11 |
| chr5_180405432_G_T | 5q35.3 | 4.1 | -3.4 | 2.3E-07 | 3% | CTD-2593A12.3/CTD-2593A12.4 |
| Chr6:30954121:G:T | 6p21.33 | 6.8 | 4.9 | 1.0E-08 | 42% | MUC21 |
| chr14_101121371_G_T | 14q32.2 | 5.8 | 3.9 | 1.0E-10 | 33% | CTD-2644I21.1/LINC00523 |
| rs12542492 | 8q21.11 | 40.0 | 2.8 | 6.3E-11 | 34% | RP11-434I12.2 |
| corect_rs116995945 | 17q22 | 3.6 | -4.5 | 2.5E-11 | 5% | SCPEP1/RNF126P1 |
| chr14_76886176_C_T | 14q24.3 | 3.5 | 2.3 | 2.3E-11 | 41% | ESRRB |
| chr2_171708059_C_T | 2q31.1 | 4.1 | -6.7 | 1.9E-11 | 7% | GAD1 |
| chr7_102368966_A_G | 7q22.1 | 4.1 | -2.6 | 6.8E-12 | 4% | RASA4DP/FAM185A |
| chr8_130380476_C_T | 8q24.21 | 4.3 | 2.5 | 4.7E-12 | 22% | CCDC26 |
| corect_rs181578054 | 22q13.33 | 4.1 | 3.0 | 7.1E-14 | 40% | ARSA/Y_RNA |
| rs3858522 | 11p15.5 | 7.7 | 3.3 | 2.2E-17 | 52% | H19/IGF2 |
| chr3_46742523_A_C | 3p21.31 | 5.2 | 4.9 | 1.8E-22 | 35% | ALS2CL/TMIE |
| chr13_113284191_C_T | 13q34 | 4.0 | -4.0 | 7.8E-23 | 5% | TUBGCP3/C13orf35 |
| chr1_97788840_A_G | 1p21.3 | 6.0 | -6.8 | 6.6E-34 | 9% | DPYD |
| chr7_118831547_C_T | 7q31.31 | 4.0 | -3.5 | 1.9E-40 | 6% | RP11-500M10.1/AC091320.2 |
| chr16_52328666_C_T | 16q12.1 | 23.0 | 5.2 | 1.5E-41 | 21% | RP11-142G1.2/TOX3 |

*Minor Allele Frequency
** < 300kb from target SNPs

**Acknowledgement**

We would like to thank the OU Supercomputing Center for Education & Research (OSCER) for supercomputing technical support, the DRIVE project for the GWAS data, NIH dbGap for data access authorization, and Dr. Xu Chao for helpful discussions. The study was funded by Dr. Pan's startup funding from the University of Oklahoma and by the Oak Ridge National Laboratory (ORNL)' Directed Research Development (LDRD) Funding. Oak Ridge National Laboratory is managed by UT-Battelle, LLC for the U.S. Department of Energy under Contract Number DE-AC05-00OR22725.

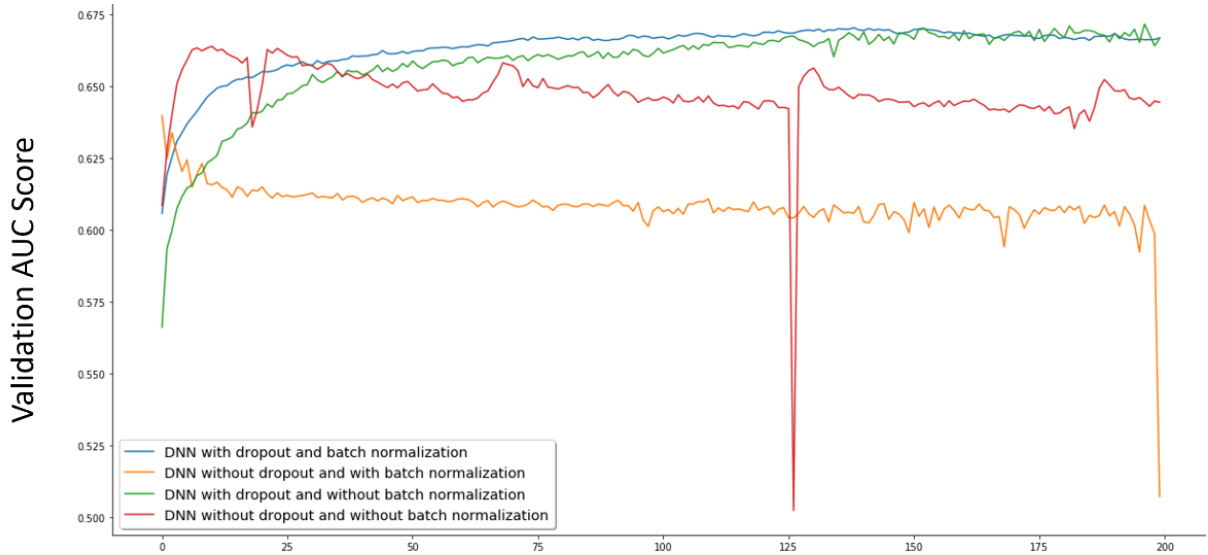

**Supplementary Figure 1: Effects of dropout and batch normalization on the 5,273-SNP DNN model.**

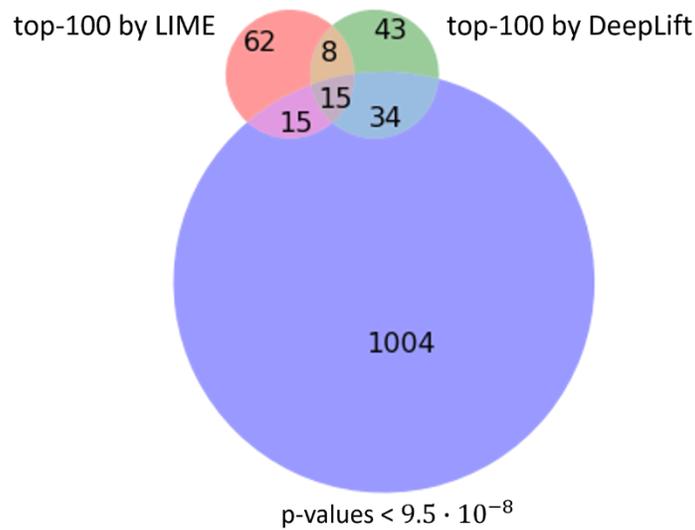

**Supplementary Figure 2: Venn diagram of important SNPs found by LIME, DeepLift, and association analysis.** The red circle represents the top-100 salient SNPs identified by LIME. The green circle represents the top-100 salient SNPs identified by DeepLift. The blue circle represents the 1,061 SNPs that had p-values lower than the Bonferroni-corrected critical value. The numbers in the Venn diagram show the sizes of the intersections and complements among the three sets of SNPs.

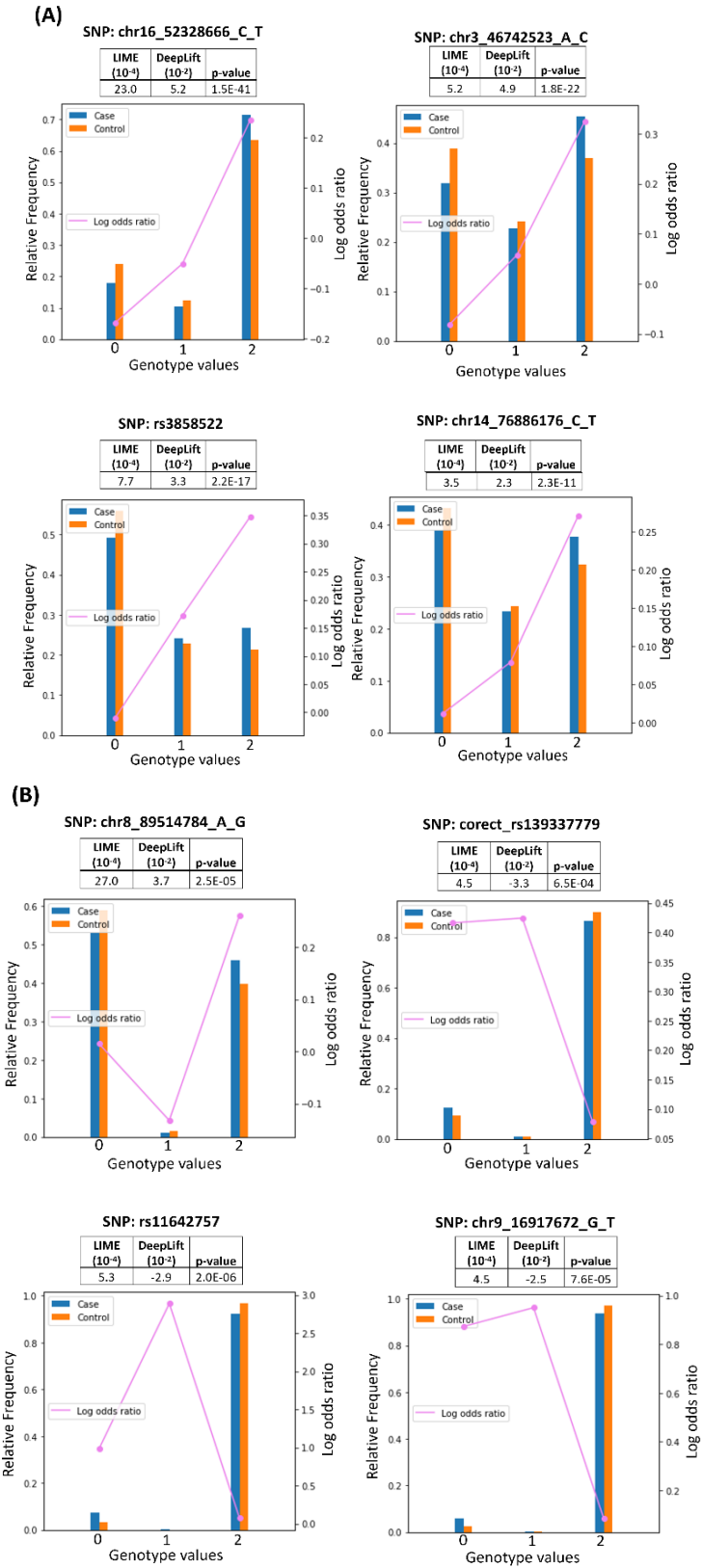

**Supplementary Figure 3: Genotype-phenotype relationships for salient SNPs used in the DNN model. (A)** Four salient SNPs with linear relationships as shown by the pink lines and the significant association p-values. **(B)** Four salient SNPs with non-linear relationships as shown by the pink lines and the insignificant association p-values. The DNN model was able to use SNPs with non-linear relationships as salient features for prediction.